\documentclass{jfm}
\usepackage{graphicx}
\usepackage{newtxtext}
\usepackage{newtxmath}
\usepackage{natbib}
\usepackage{hyperref}
\usepackage{gensymb}
\usepackage{tikz}
\usepackage{subcaption}
\hypersetup{
    colorlinks = true,
    urlcolor   = blue,
    citecolor  = blue,
}

\newcommand{\RomanNumeralCaps}[1]
\usepackage{verbatim}

\title{Existence of three distinct scaling regimes in self-propelled airfoil}

\author{Rakshita Joshi\aff{1}
  \corresp{\email{rakshitaj@iisc.ac.in}},
 \and Jaywant Arakeri\aff{1}}

\affiliation{\aff{1}Indian Institute of Science, Bangalore, India}

\begin{document}
\maketitle

\begin{abstract}
We investigate the effect of imposed kinematics on the self-propulsion of the NACA0015 symmetric airfoil section subject to sinusoidal pitching. We employ a rotary apparatus capable of achieving self-propulsion. A power-spring-based crank-rocker mechanism actuates the airfoil. Three distinct scaling relations emerge, which relate the self-propulsion Reynolds number $Re_s$ to the frequency Reynolds number $Re_f$, the amplitude of pitching $\theta_0$, and the location of the pitching point, $p$. When pitched near the center, a \textit{linear} scaling emerges with $Re_s \sim Re_f \theta_0$. When pitched near the leading edge, a \textit{power} scaling emerges with $Re_s \sim (1-2p)(Re_f \theta_0)^{3/2}$ for low amplitude pitching and a \textit{separable} scaling emerges with $Re_s \sim (1-2p)^{1/2}Re_f\theta_0^{1/2}$ for moderate to high amplitude pitching. These relations are consistent with the scaling relations derived from balancing inviscid thrust with viscous drag, pressure drag, and enhanced pressure drag for the \textit{power}, \textit{separable}, and \textit{linear} regimes, respectively. We find that different vortical patterns in the wake are directly correlated to the airfoil's self-propulsion speed which essentially determines the spatial separation between the shed vortices. Our findings provide a comprehensive framework for understanding the self-propulsion of rigid pitching airfoils across a wide range of parameters validated experimentally. 
\end{abstract}


\section{Introduction}
\label{intro}
\begin{comment}
    Zoologist Gray, in an effort to determine the propulsive power of cruising dolphins \citep{gray1936}, stumbled upon a paradox – the drag power of the dolphins exceeds its muscle power. Although it's now considered resolved (see, \citet{fish2006} and \citet{bale2015}), Gray's paradox had long perplexed researchers studying the hydrodynamics and energetics of swimming fishes – with a singular interest in the mechanics of thrust generation and swimming efficiency. The resolution of the paradox revealed a fundamental aspect of fish propulsion – muscle energy is used to produce lateral undulations generating thrust, which balances the drag. 
\end{comment}
For a fish cruising at a constant speed, the net force experienced by it must be zero and, therefore, is said to be "self-propelling". Self-propulsion ensues when prescribed actuation generates sufficient thrust to overcome the hydrodynamic resistance, establishing a precise balance between thrust generated and drag experienced. Consequently, for self-propelling bodies, a crucial and foundational inquiry arises: what factors determine swimming speed, and how are they influenced by imposed kinematics such as frequency, amplitude, or the mode of oscillation?

Oscillating airfoil exhibit a remarkable ability to self-propel, rendering itself as simple model objects for such investigations. See, for example, \citet{lauder2007}, \citet{alben2012}, \citet{gazzola2014}, \citet{das2016,das2019,das2022}, \citet{lagopoulos2019}, \citet{liu2020}, \citet{wang2020}, \citet{gross2021}, \citet{paniccia2021}, \citet{lin2021} and \citet{wu2022}. Although flexibility is ubiquitous and a crucial parameter in fish swimming, the unsteady aerodynamics of rigid airfoils are much better understood and boasts a strong theoretical background.  By decoupling the effect of flexibility, we may better understand the fundamentals of the self-propulsion mechanism. The effect of flexibility may then be interpreted and understood as a deviation from rigid behavior.

The spontaneous forward motion of a heaving flat plate and elliptical airfoils reported first by \citet{vandenberghe2004} and later by \citet{ spagnolie2010} marks an early and systematic exploration of self-propulsion in oscillating airfoils. In both studies, the airfoil propulsion occurs beyond a critical frequency, and thereafter, the propulsion speed increases linearly with the frequency. On the contrary, numerical simulations of heaving elliptical airfoils \citep{alben2005, zhang2010, wang2022} report a non-linear relationship between self-propelling speed and heaving frequency. However, the self-propulsion Reynolds number is $O(10^2)$ in numerical simulations, an order of magnitude lower than those studied in the experiments cited earlier. Experimental studies on pitching symmetric airfoils are also very limited. Notable contributions include those of \citet{lauder2007} - on the self-propulsion characteristics of two tandem flapping airfoils - and \citet{mackowski2015} - report a drag-to-thrust transition when pitching at a small amplitude of 2°. Although in recent years, there are a growing number of numerical simulations of pitching symmetric airfoil in a 2-dimensional flow \citep{das2016,das2019,das2022,lin2021,chao_tailbeat_2024}, experimental studies are scarce. 

An important result in the numerical studies on self-propulsion in pitching airfoils is the emergence of a scaling relation between the self-propelling Reynolds number $Re_s = U_sC/\nu$ and a trailing edge Reynolds number $Re_{TE} = fAC/\nu$ (also called "flapping Reynolds number") of the form $Re_s \sim Re_{TE}^\alpha$. Here, $U_s$ is the average self-propulsion speed, $C$ is the airfoil chord length, $A$ is the total trailing edge excursion and $\nu$, the kinematic viscosity of the fluid medium. Since $A$ is related to $\theta_0$ and pitching location $p$ i.e., $A = 2\sin{\theta_0}(1-p)C$, $Re_{TE}$ is expected to account for all the imposed kinematic parameters. \citet{das2016, das2019, das2022} and \citet{lin2021} report a value of $\alpha \approx 5/3$ valid up to $Re_s \sim O(10^3)$ and $Re_s \sim O(10^2)$ respectively. \citet{chao_tailbeat_2024} report $\alpha \approx 4/3$ valid up to $Re_s \sim O(10^2)$. A similar scaling law for macroscopic elastic swimmers with $\alpha \approx 4/3$ was reported by \citet{gazzola2014} for $Re_s < O(10^3)$. Note that the "swimming number" used by \citet{gazzola2014} has the same definition as $Re_{TE}$ except that the trailing edge excursion $A$ is a result of elastic movement of the swimmers and not an imposed kinematic condition as is the case in rigid airfoils. They arrive at these scaling relations by considering a dynamic balance between the average reactive force and skin friction \citep{gazzola2014, lin2021, chao_tailbeat_2024} or "enhanced skin friction" \citep{das2016,das2019,das2022}. The varying $\alpha$ values in the scaling relations arise from different assumptions about the drag forces. While these scaling relations are consistent in capturing the effect of frequency on self-propulsion speed, they are, however, inconsistent in accounting for the effect of amplitude. Specifically, \citet{das2022} note the amplitude dependence of the proportionality constant (prominent around $Re_s \sim O(10^3)$) of the scaling relations. 

Another parameter affecting the propulsive characteristics of a pitching airfoil is the location of the pitching point along the chord. This location has a direct bearing on the net reactive forces and the development of leading-edge vortices on the airfoil. In the context of thrust generation, both \citet{mackowski_effect_2017} and \citet{tian_experimental_2016} observe that the net thrust coefficient decreases as the pitching point is shifted from the leading edge to the trailing edge. For self-propelling airfoils, \citet{LIN201933} find that the self-propulsion speed monotonically decreases with the pitch point location shifted from the leading edge to the trailing edge. \citet{das2022}, on the other hand, consider a case where the airfoil is "bi-directionally free" i.e., passive heave being permitted. With this passive heave, the "effective pitching point" location is shifted to the center of the chord. For such a configuration, the self-propulsion speed is lower than the pure-pitching case. Similar to amplitude dependence, the proportionality constant in the $Re_s - Re_{TE}$ relation is adjusted to account for the change in effective pitching point.

Despite the valuable insights gained from numerical investigations, there remains a prominent scarcity of experimental findings to validate the numerical results and address the discrepancies arising, particularly in varying values of the scaling exponent $\alpha$ and the inability of these relations to effectively capture the effect of amplitude and pitching point location through $Re_{TE}$. To address these concerns, we experimentally investigated the self-propulsion of a sinusoidally pitching airfoil. Our work provides the much-needed experimental database for this problem and addresses the discrepancies in accounting for the effects of fundamental kinematic parameters. Specifically, we examine the effect of pitching frequency ($f$), amplitude ($\theta_0$), and the location of pitching point ($p$). We identify not one but three distinct scaling regimes –{\it power, separable, and linear scaling regimes} – relating the self-propulsion Reynolds number $Re_s$ to the frequency Reynolds number ($Re_f = (fC)C/\nu$), $\theta_0$ and $p$. We highlight the different mechanisms of invisicd thrust generation and the effect of pitching point point locations in these mechanisms. The three distinct relations emerge when we take into account different drag contributions - enhanced skin friction, pressure drag, and an enhanced pressure drag - balancing the inviscid thrust. A major contribution of this work is the identification of the \textit{separable} scaling and \textit{linear} scaling, which has not been reported earlier. By using three separate non-dimensional parameters instead of a single parameter $Re_{TE}$, these scaling relations more clearly highlight the effect of the individual parameters. Here, the \(Re_s\) values range from \(O(10^2)\) to \(O(10^4)\), with the upper limit being an order of magnitude higher than that reported in previous studies. We also find a strong correlation between the self-propulsion speed and vortex patterns shed in the wake. To the best of our knowledge, this is the first experimental work to systematically explore all three pitching parameters and explicitly establish their individual effects on self-propulsion speed. 

The article is structured as follows: section \ref{method} provides an overview of the experimental method; section \ref{speed} presents the measured self-propelling speed and explores the impact of imposed kinematics; section \ref{scaling}, presents the scaling relations; section \ref{wake} presents wake vortex patterns; section \ref{concl} concludes the article while summarizing key findings.  

%----------------------------------------------------------------------------------------------------------
\section{Experimental Methodology}
\label{method}
\begin{figure} % 'h' stands for 'here'; you can use other placement options such as 't' for top, 'b' for bottom, etc.
\centerline{\includegraphics[width = 1\textwidth]{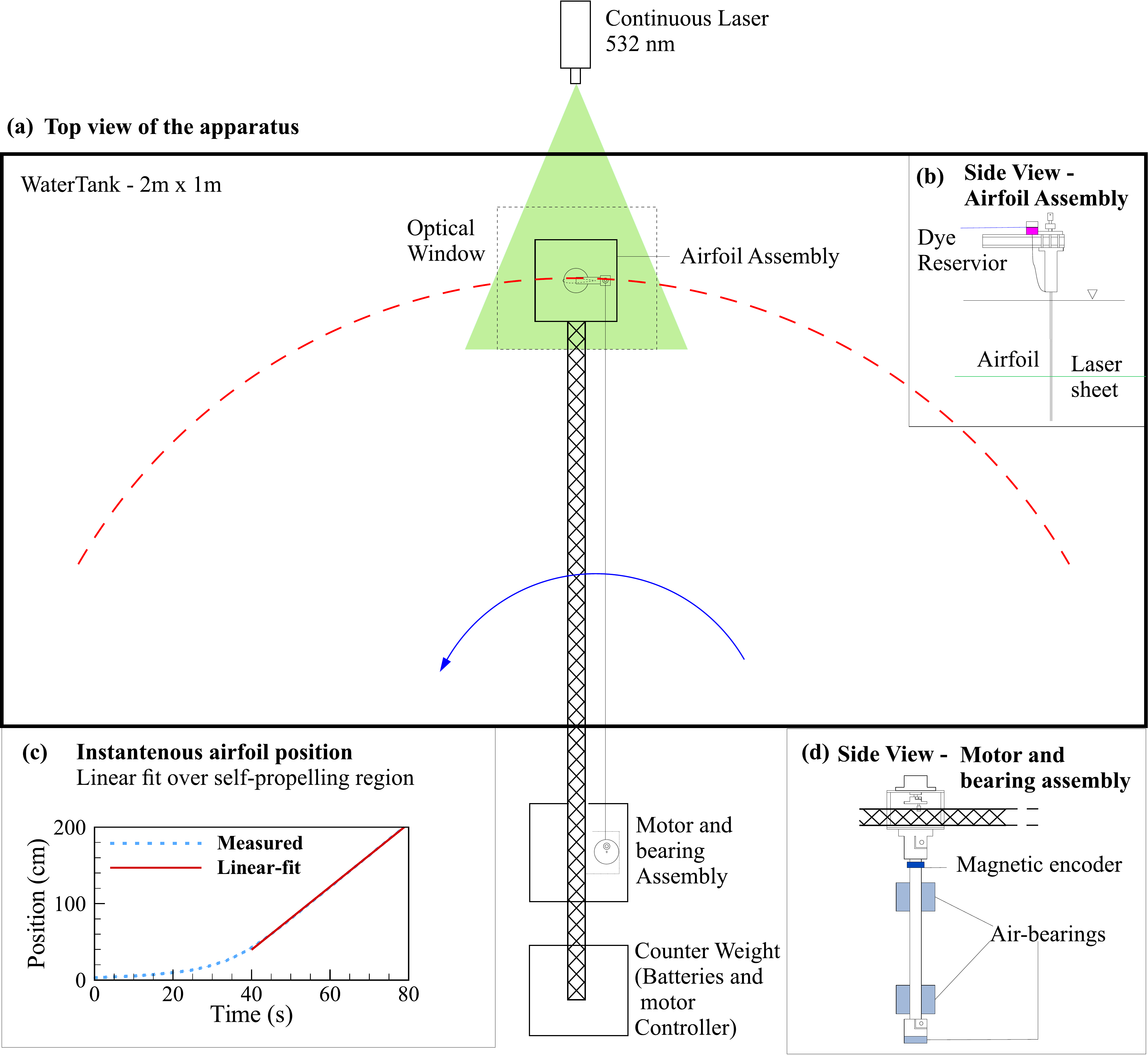}}% Adjust the width as needed
\caption{(a) Schematic of the top view of the apparatus. The red-dashed line indicates the path traversed by the airfoil. (b) Side view of the airfoil assembly. (c) The position-time graph shows the instantaneous position of the airfoil with time for pitching at $2\,\text{Hz}$ with an amplitude of $9.5\degree$. The slope of the linear fit is the self-propelling velocity. (d) Side view of the motor and bearing assembly showing the freely rotating shaft supported by air-bearings.}
\label{schematic}
\end{figure}

 We use a NACA0015 airfoil section with chord length $C = 38.9 \, \text{mm}$ and span $h = 100.1\, \text{mm}$. The airfoil is made with Acrylonytril Butadyne Sulphide (ABS) by vacuum casting. The density of the airfoil is $1.01 \, \text{g/cm}^3$, ensuring the airfoil is close to being neutrally buoyant. Details of the airfoil fabrication process can be found in \citet{shinde2012}. The airfoil propels in a large water tank measuring $2\,\text{m} \times 1\,\text{m} \times 0.5\,\text{m}$. We use a rotary apparatus, similar to the "merry-go-round" set-up of \citet{thiria2010}, to achieve self-propulsion. Figure\ref{schematic} shows the schematic of the top view of the apparatus. The airfoil is positioned at the end of a long, freely rotating arm with an effective length of $1.04\,\text{m} (>20C)$. The arm is allowed to rotate for about $120 \degree$, resulting in a path length of about $2\,\text{m}$ even after allowing for wall effects. The total path length is more than the linear distance between the two endpoints due to the curved path of the airfoil. Nevertheless, since the airfoil dimensions are much smaller than the arm length, curvature effects on the self-propulsion can be neglected in such a configuration.
 
 The rotating arm is essentially a truss structure fabricated in-house with carbon fiber. This truss (rotating arm) is securely mounted on a high-precision steel shaft. The shaft is supported by a combination of axial and thrust (flat) air bearings (figure\ref{schematic}). The bearings allow free rotation of the shaft and the truss while constraining motion in other directions. The use of air bearings ensures that the friction from the supporting structure is negligible. A DC motor is mounted near the shaft while the batteries required to power the motor and the controllers are mounted as a counterweight (figure\ref{schematic}). A spiral-spring-based crank rocker mechanism is used to convert the rotatory motion into sinusoidal oscillation and transmit it at a large distance. The rocker of the mechanism is connected to the airfoil through a spiral power spring. $M3 \times 20\,\text{mm}$ threaded holes are positioned along the chord line to allow for adjustments in the pitching point location. The airfoil is connected to the spiral spring through a pair of ball bearings. This connecting shaft is of diameter $3\,\text{mm}$. The airfoil is fully submerged at a depth of $5\,\text{cm}$ from the surface.
 
 We use an incremental magnetic rotary encoder (Reineshaw\copyright LM10) with an angular resolution of $0.018\degree$ to measure the instantaneous position of the shaft. The encoder is directly mounted on the shaft itself (see figure\ref{schematic}). The linear resolution of the position measured at the airfoil end is $0.3\,\text{mm}$ ($\approx0.007C$). The position-time plot in figure\ref{schematic} shows the typical measured linear position with time for a representative case. The steady self-propelling state of the airfoil is identified as a linear region in the position-time curve. A linear function is fitted to the data within this region. The slope of this linear segment averaged across $120-150$ oscillation cycles (in $10-15$ trials), is reported as the self-propelling speed $U_s$. The nature of this position-time curve is similar for all the cases except for variations in the time taken to achieve self-propulsion and the slope of the linear region. A more detailed discussion about the apparatus, particularly focused on the kinematics of the mechanism and characterization of self-propulsion, can be found in \citet{joshi2022, joshi2024}. 
 %-----------------figure- Schematic----------------
%-----------------------------------------

\begin{figure}% 'h' stands for 'here'; you can use other placement options such as 't' for top, 'b' for bottom, etc.
\centerline{\includegraphics[width = 0.7\textwidth]{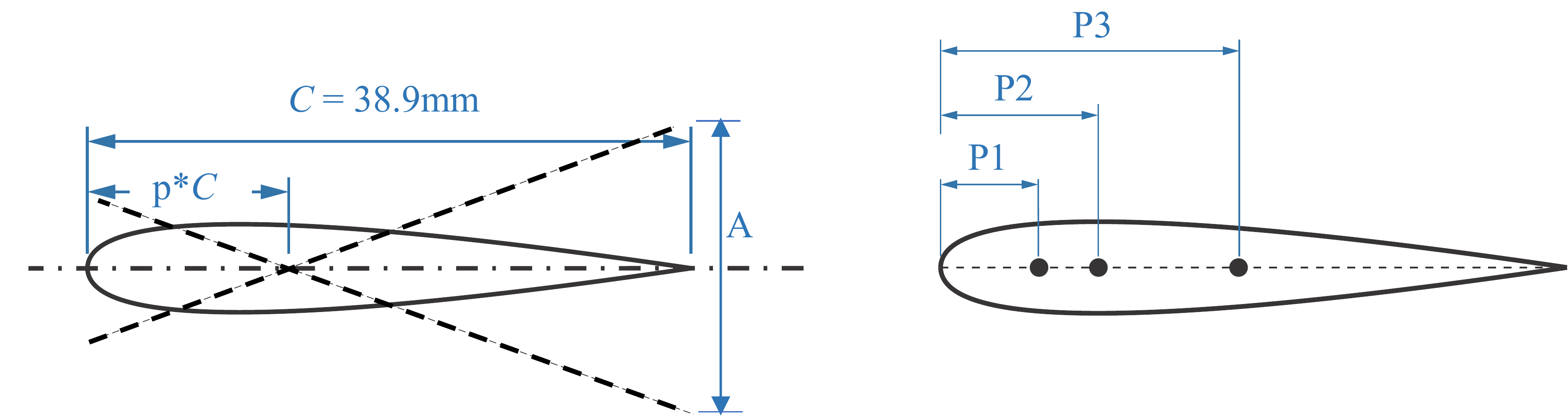}} % Adjust the width as needed
\caption{Illustration of the NACA0015 airfoil section with depictions of the trailing edge excursion on the left and the locations of pitching points on the right. The pitching location P1, P2 and P3 are at $p = 0.125$, $p = 0.3$ and $p - 0.48$ respectively.}
\label{airfoil}
\end{figure}

We pitch the airfoil sinusoidally such that the instantaneous pitch angle of the airfoil is given by $\theta=\theta_0 \sin(2\pi ft)$. We consider four different amplitude values - $\theta_1 \approx 4.6\degree$, $\theta_2 \approx 9.4\degree$, $\theta_3 \approx 14.8\degree$ and $\theta_4 \approx 20.5\degree$. The location of pitching point $p$ represents the location as a fraction of the airfoil chord. As such, $p = 0$ at the leading edge and $p = 1$ at the trailing edge. Consequently, when pitched at $p$, the actual distance of the pitching point is $pC$ from the leading edge. In the present work, we consider three pitching points - P1, P2, and P3 - at $p = 0.125, 0.3$ and $0.48$ respectively. All the geometric parameters and the pitching point locations are sketched in figure\ref{airfoil}. The pitching frequency varies from \(0.25 \, \text{Hz}\) to \(8 \, \text{Hz}\), depending on the amplitude and pitching point. For a given amplitude and pitch location, the lowest frequency is the one where self-propulsion is reliably achieved and consistently repeatable across the trials. The largest frequency is limited by the occurrence of water sloshing and/or transverse (span-wise) oscillations of the truss. The self-propulsion speed ranges from approximately \(2 \, \text{cm/s}\) to \(45 \, \text{cm/s}\), with the corresponding \(Re_s\) values ranging from $400$ to $17,000$, spanning two orders of magnitude. 

We use Laser-Induced Fluorescent (LIF) technique to visualize the flow qualitatively and observe the vortex patterns in the wake. A 532nm laser sheet illuminates the mid-span of the airfoil, positioned near the center of the water tank. At this position the airfoil is already in a self-propelling state (figure \ref{schematic}). Rhodamine B dye is used in varying concentrations depending on the propulsion speed. The dye is continuously injected from a small reservoir (with a 5ml capacity) near the airfoil end. A thin tube connects the reservoir to the dye injection port on the airfoil located at $p = 0.2$. This port opens at the mid-span, releasing dye at the airfoil surface through a porous plug, ensuring smooth release of dye. The dye flow rate is adjusted for each propulsion speed and tested by towing the airfoil at different speeds. Instantaneous images of the flow field are captured using a high-speed camera. 

%------------figure- Airfoil Schematic-----------
%-------------------------------------------------------------
\section{Self-propulsion Speed}
\label{speed}
\begin{figure}
    \centering
    \begin{subfigure}{0.555\textwidth}
        \includegraphics[width = \textwidth]{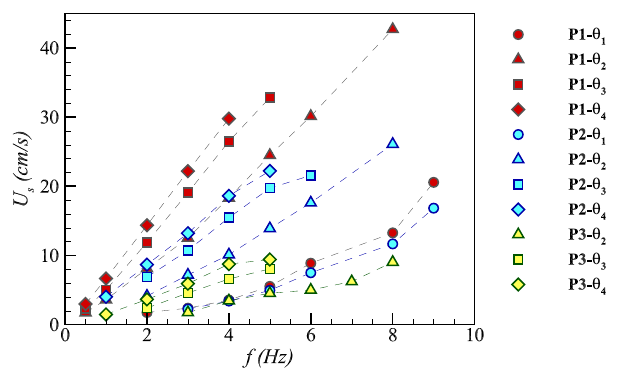}
        \caption{}
        \label{u:f}
    \end{subfigure}
    \begin{subfigure}{0.42\textwidth}
        \includegraphics[width = \textwidth]{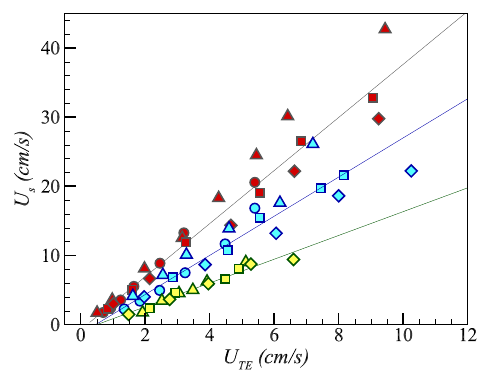}
        \caption{}
        \label{U:UTE}
    \end{subfigure}
    \caption{Variation of the self-propelling velocity $U_s$ with $f$ (left) and  trailing edge velocity scale $U_{TE} = fA$ (right) for all the parameters studied. The dashed lines (left) represent a piece-wise linear interpolation. Solid lines (right) represent the best fit.}
    \label{U_s}
\end{figure}

In figure \ref{U_s}, we plot $U_s$ with frequency for all amplitude and pitching points and $U_s$ with trailing edge velocity, $U_{TE}$, for all the parameters. The error in measurement, defined as twice the standard deviation - is less than the symbol size. Note that we do not observe any self-propulsion when pitched at P3 with amplitude $\theta_1$.Recall from section\ref{intro} the use of $Re_{TE}$ as a parameter used to determine the scaling relations in self-propulsion. In the expression $Re_{TE} = fA C / \nu$, the term $fA$ essentially provides a trailing edge velocity scale. Therefore, we can rewrite, $Re_{TE}$ as $Re_{TE} = U_{TE} C / \nu$, where $U_{TE} = fA$. This definition of $U_{TE}$ accounts for all three pitching parameters since $A  = 2\sin{\theta_0}(1-p)C$.  

For a given pitching point, $U_s$ increases with frequency and amplitude (figure \ref{u:f}). $U_s$ value ranges from about $2 \,\text{cm/s}$ to nearly $45\,\text{cm/s}$ for pitching location P1, $26\,\text{cm/s}$ for P2 and $10\,\text{cm/s}$ for P3. In general, we observe larger $U_s$ values for P1, with $U_s$ decreasing as the pitching point is moved closer to the center (P3). In figure \ref{U:UTE}, $U_s$ still exhibits amplitude and pitching point dependence for pitching locations P1 and P2 even when plotted with $U_{TE}$ which is expected to account for the variation in amplitude and the pitching point location. Only for P3, We see a linear increase in $U_s$ with $U_{TE}$ with all the points collapsing on a straight line. Therefore, $U_{TE}$, although a function of $\theta_0$ and $p$, cannot adequately capture their individual effects on self-propulsion for pitching locations, especially at P1 and P2. In subsequent sections, we will discuss how individual effects can be explicitly explored without the use of $U_{TE}$ (and subsequently $Re_{TE}$) as an input velocity scale. 

We represent self-propulsion speed in terms of another non-dimensional parameter $U^*_{BL} = U_s/fC$. $U^*_{BL}$ represents the speed in terms of body lengths per oscillation; a commonly used non-dimensional parameter to compare the swimming speed of different fishes \citep{videler1993}. The examination of steady-state swimming in oscillatory fishes from different species reveals that the normalized speed $U^*_{BL}$ ranges between $0.2$ and $0.8$ \citep{videler1993}. Additionally, their tail-beat excursion typically falls within the range of $10 - 30\%$ of their body length \citep{videler1993}. Similar values of trailing edge excursion are achieved for rigid pitching airfoil when pitching at $\theta_2 \approx 9.4 \degree$. By definition, $U^*_{BL}$ is related to the inverse of the self-propelling reduced frequency $k_s$ such that $U^*_{BL}={\pi}/{k_s}$. 

\begin{figure}
    \centering
    \begin{subfigure}{0.555\textwidth}
        \includegraphics[width = \textwidth]{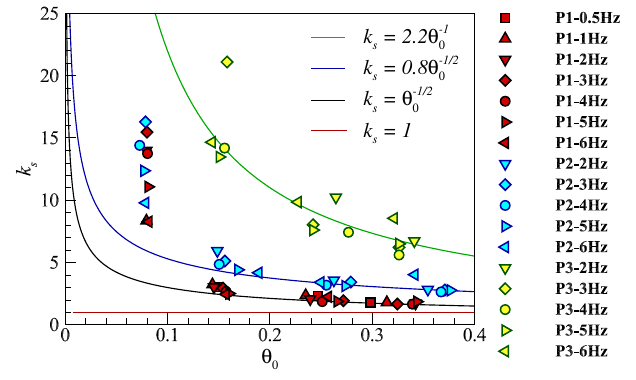}
        \caption{}
        \label{k:th}
    \end{subfigure}
    \begin{subfigure}{0.42\textwidth}
        \includegraphics[width = \textwidth]{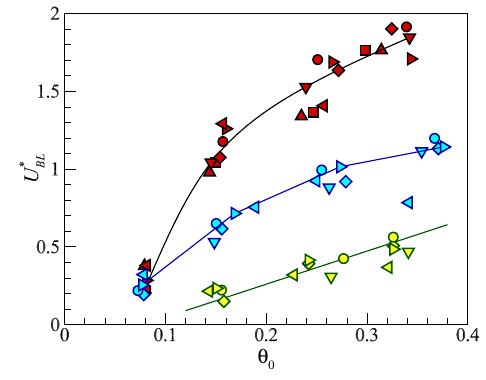}
        \caption{}
        \label{ubl:th}
    \end{subfigure}
    \caption{Self-propulsion reduced frequency $k_s = \pi fC/U_s$ (left) and non-dimensional speed $U^*_{BL} = U_s/fC$ (right) with pitching amplitude $\theta_0$ in radian. Note that $U^*_{BL} = \pi/k_s$.}
    \label{k:ubl}
\end{figure}

In figure \ref{k:th} and figure \ref{ubl:th} we plot $k_s$ and $U^*_{BL}$ with pitching amplitude $\theta_0$. Note that unless explicitly specified, $\theta_0$ values are considered in radian for all calculations. For the range of parameters considered here, we observe a wide range of $k_s$ values, from $1$ to $22$, indicating a considerable variation in the degree of unsteadiness across all examined parameters. However, this variation is prominent for pitching location P3 in partiular. For pitching locations P1 and P2, the variation in $k_s$ is prominent only at lower pitching amplitude ($\theta_1$, corresponding to $\theta_0 <0.1$rad). For pitching amplitudes $\theta_2$, $\theta_3$ and $\theta_4$ (essentially for $\theta_0 > 0.14$rad) the variation of $k_s$ with $\theta_0$ is well captured by the relation $k_s \sim \theta_0^{-1/2}$ for pitching locations P1 and P2 and $k_s \sim \theta_0^{-1}$ for P3. Here, points corresponding to different frequencies cluster closely together for nearly the same $\theta_0$ indicating a linear dependence of self-propulsion speed on frequency. In figure \ref{ubl:th}, notice that the data points for each pitching point are along a single curve and $U^*_{BL}$ increases as the pitching point approaches the leading edge i.e., $p$ decreases. The maximum value of $U^*_{BL}\approx 2$ is noted when pitching closest to the leading edge (P1 with $p = 0.125$) for pitching amplitude $\theta_4$. Generally, $U^*_{BL}$ increases with $\theta_0$. However, for pitching points P1 and P2 and $\theta_0 > 0.14$rad, the rate at which $U^*_{BL}$ increases with $\theta_0$ reduces, indicating reduced sensitivity to further increase in amplitude. This diminishing effect of amplitude is more prominent in the $k_s$ vs $\theta_0$ plot (figure \ref{k:th}).

%-------------------------------------
\section{Scaling regimes}
\label{scaling}
To independently assess the effect of all the pitching parameters, we define a frequency Reynolds number $Re_f = (fC)C/\nu$ where the longitudinal velocity scale $fC$ dependents only on the pitching frequency and explore the effect of $Re_f$, $\theta_0$ and $p$ on $Re_s$. We will discuss how the three scaling regimes emerge, the differing thrust and drag mechanisms and validate these relations with our experimental data.  
\subsection{Thrust generation as an inviscid phenomenon}
\label{thrust_gen_in}
For a thrust-generating airfoil oscillating in a uniform free stream, the instantaneous net propulsive thrust coefficient $C_T$ would consist of an average and an unsteady component. 
\begin{equation}
    C_T = \bar{C_T} + C_{T0}\sin{(2\omega t + \phi)}
\end{equation}
Here, $\bar{C_T}$ is the average thrust coefficient, $C_{T0}$ is the amplitude of the unsteady component, $\omega = 2\pi f$ is the angular frequency and $\phi$ is the phase difference between the propulsive thrust and the sinusoidal oscillation of the airfoil. \citet{mackowski2015} found that the directly measured $C_{T0}$ and $\phi$ values matched very well with the predicted thrust values from the linear inviscid theory of \citet{garrick1936}. Their finding established that the thrust generated by a pitching airfoil is primarily due to inviscid mechanisms well captured by the linear inviscid theory of \citet{theodorsen1935} and \citet{garrick1936}. An important consequence of these findings is that the average thrust $\bar{C_T}$ can now be determined by offsetting the inviscid average thrust $\bar{C}_{T,in}$ with a suitable average drag coefficient $\bar{C}_D$ i.e.,
\begin{equation}
    \bar{C_T} = \bar{C}_{T,in} - \bar{C}_D
    \label{thrust_net}
\end{equation}
In a self-propelling state, $\bar{C}_T = 0$. Consequently, from Eq. \ref{thrust_net} the invisicd thrust generated by the airfoil balances the drag forces and $\bar{C}_D = \bar{C}_{T,in}$.

The linear inviscid thrust is essentially a linear superposition of the reactive thrust ($\bar{C}_R$), circulatory thrust ($\bar{C}_C$), including the quasi-steady circulation due to non-zero angle of attack, and the induced vorticity due to the wake, and leading-edge suction ($\bar{C}_{Les}$). As such, we can write,
\begin{equation}
    \bar{C}_{T,in} = \bar{C}_R + \bar{C}_C + \bar{C}_{LeS}
\end{equation}

In  Appendix \ref{invisicdthrust}, we revisit the invisicd theory \citep{garrick1936} to derive the expressions for $\bar{C}_R,\bar{C}_C,\bar{C}_{LeS}$ and $\bar{C}_{T,in}$ as function of reduced frequency $k = \pi fC/U$ for a pitching flat-plate airfoil in uniform free-stream. In the limit of $k \rightarrow \infty$, we see that $\bar{C}_{T,C} \approx 0$ while
\begin{equation}
    \bar{C}_R = \pi \theta_0^2\frac{k^2}{2}\left(1-2p\right)
%    \label{Ct:Reac}
\end{equation}
and
\begin{equation}
    \bar{C}_{LeS} \approx \pi \theta_0^2 \frac{k^2}{4} \left( 2p - \frac{1}{2} \right)^2
    \label{Ct:Les}
\end{equation}
with the total invisicd thrust
\begin{equation}
    \bar{C}_{T,in} \approx \pi\theta_0^2\frac{k^2}{4}\left(\frac{3}{2} -2p\right)^2
\end{equation}
Since all the thrust coefficients scale as $\pi\theta_0^2k^2/2$, we asses the effect of pitching point location on normalized thrust coefficients $\bar{C}_R^* = \bar{C}_R/(\pi\theta_0^2k^2/2)$, $\bar{C}_{LeS}^* = \bar{C}_{LeS}/(\pi\theta_0^2k^2/2)$ and $\bar{C}_{T,in}^* = \bar{C}_{T,in}/(\pi\theta_0^2k^2/2)$. 

\begin{figure}
    \centering
    \includegraphics[width = 0.75\textwidth]{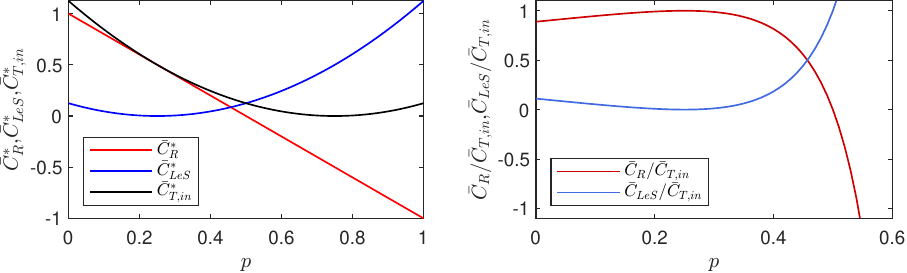}
    \caption{Normalised inviscid thrust coefficients and their relative contribution to the total thrust in the asymptotic limit of $k \rightarrow \infty$.}
    \label{Ct:p}
\end{figure}

In figure \ref{Ct:p} we plot the variation of $\bar{C}_R^*$, $\bar{C}_{LeS}^*$, $\bar{C}_{T,in}^*$ with the pitching point location. We also plot the variation of fractional contribution $ \bar{C}_R/\bar{C}_{T,in}$ and $ \bar{C}_{LeS}/\bar{C}_{T,in}$ with pitching point location. The reactive thrust contribution is such that it generates a positive thrust force when $p<1/2$ and a negative thrust when $p>1/2$ with $\bar{C}_R^*$ linearly decreasing with $p$. At $p = 1/2$, $\bar{C}_R = 0$. The contribution of leading edge suction to the total invisicd thrust is zero when pitched at $p = 1/4$, but positive at all other locations. We see larger values of $\bar{C}_{LeS}^*$ at the trailing edge side of the airfoil beyond $p>1/2$. At three-quarter chord i.e., $p = 3/4$, the negative reactive thrust is balanced by the positive thrust due to leading edge suctions resulting in net zero average thrust. Nevertheless, for $p<1/2$, $\bar{C}_{LeS}^*$ values are much smaller compared to $\bar{C}_R^*$. The total thrust is dominated by reactive forces until $p\approx 0.4$, evident from the variation of $\bar{C}_R/\bar{C}_{T,in}$ with $p$ in figure \ref{Ct:p}. Therefore, for pitching locations with $p<0.4$, we expect the total inviscid thrust to scale as the reactive thrust itself such that 
\begin{equation}
    \bar{C}_{T,in} \sim \bar{C}_R \sim \theta_0^2 k^2(1-2p)
    \label{Ct:R}
\end{equation}
However, when pitched close to the mid-chord ($p \approx 1/2$), the thrust generation would be primarily due to leading edge suction such that
\begin{equation}
    \bar{C}_{T,in} \sim \bar{C}_{LeS} \sim \theta_0^2k^2
    \label{Ct:LeS}
\end{equation}

\begin{figure}
\centering
  \begin{subfigure}{0.3\textwidth}
    \centering
    \includegraphics[width=1.0\linewidth]{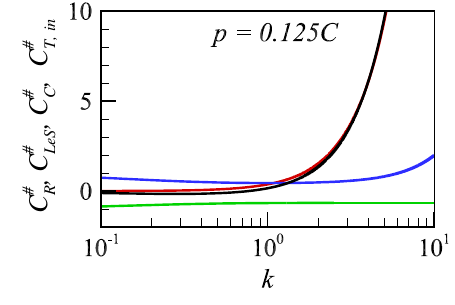}
  \end{subfigure}
%\hfill
  \begin{subfigure}{0.3\textwidth}
    \centering
    \includegraphics[width=1.0\linewidth]{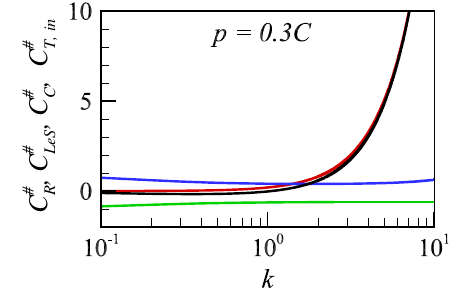}
  \end{subfigure}
%\vfill
  \begin{subfigure}{0.3\textwidth}
    \centering
    \includegraphics[width=1.0\linewidth]{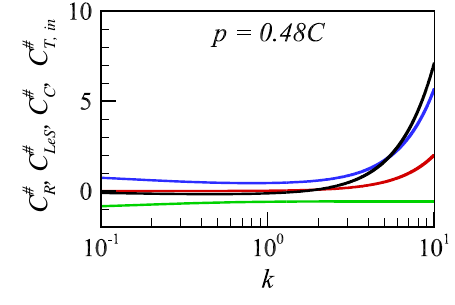}
  \end{subfigure}
  
  \caption{The components of the total average inviscid thrust coefficient (in black): reactive forces (in red), circulatory lift component (in green), and leading-edge suction (in blue).}
\label{Ct:ks}
\end{figure}

In figure \ref{Ct:ks} we compare the variation of $\bar{C}_R,\bar{C}_C$, $\bar{C}_{LeS}$ and $\bar{C}_{T,in}$, normalised by $\pi \theta_0^2$, with $k$ for pitching locations P1 ($p = 0.125$), P2 ($p = 0.3$) and P3 ($p = 0.48$). Note that $C^\#_R = \bar{C}_R/\pi\theta_0^2$,$C^\#_C = \bar{C}_C/\pi\theta_0^2$, $C^\#_{LeS} = \bar{C}_{LeS}/\pi\theta_0^2$ and, $C^\#_{T,in} = \bar{C}_{T,in}/\pi\theta_0^2$. We see that the asymptotic behavior of $\bar{C}_{T,in}$ from Eq.\ref{Ct:R} (for P1 and P2) and Eq.\ref{Ct:LeS} (for P3) derived in the limit of $k \rightarrow \infty$ is valid for $k>1$ and $k >5$ respectively. Recall from our earlier discussions and figure \ref{k:ubl} that the measured $k_s$ values are within the range of validity of these asymptotic relations. 

\subsection {Power scaling regime}
\label{psr}
When pitching at low amplitude ($\theta_1$), viscosity is expected to influence self-propulsion and the impact of skin friction on self-propulsion would be significant. Moreover, the presence of a moving airfoil surface further enhances skin friction in accordance with the Bone-Lighthill boundary layer thinning hypothesis \citep{lighthill1971}. While the boundary layer skin friction scales as $Re^{-1/2}$ for laminar flow over stationary boundary, it increases as $Re^{-1/2}|U^*_n|^{1/2}$ for moving surfaces \citep{ehrenstein2014}, where $U^*_n$ represents a non-dimensional normal velocity. For a pitching airfoil, $U^*_n \sim k_s\theta_0$. Consequently, for the self-propelling airfoil, the cycle-averaged drag coefficient $\bar{C_D}$ would then scale as
\begin{equation}
\bar{C_D} \sim Re_s^{-1/2}(k_s \theta_0)^{1/2}
\label{CDviscous}
\end{equation}
From Eq.\ref{Ct:R} and Eq.\ref{CDviscous} we get 
\begin{equation}
Re_s^{-1/2} \sim (1 -2p)k_s^{3/2}\theta_0^{3/2}
\label{ps1}
\end{equation}
and upon rearranging, 
\begin{equation}
Re_s \sim (1-2p) (Re_f\theta_0)^{3/2}
\label{psmain}
\end{equation}
In figure \ref{Re:Pow}, we compare our experimental data and find a close agreement with the predicted scaling for pitching locations P1 and P2 at low amplitude $\theta_1$ up to $Re_f\theta_0 \approx 2500$, the tested limit in this study. We also compare our findings with those reported by \citet{mackowski2015} and \citet{das2016}. Notice that all the data points fall within the curve defined by $Re_s \sim (1-2p) (Re_f\theta_0)^\alpha$ with $\alpha$ varying between $4/3$ \citep{gazzola2014} and $5/3$\citep{das2016, lin2021}. If one assumes that dominant resistance is only due to viscous skin friction \citep{gazzola2014}, then $\bar C_D \sim Re_s^{-1/2}$ alone, resulting in $\alpha = 4/3$. A closer examination of figure \ref{Re:Pow} shows that the exponent $\alpha = 4/3$ might be more appropriate for lower $Re_f\theta_0$ where the boundary layer thinning may not be very prominent. While at higher $Re_f\theta_0$ values, enhancement in skin friction is considerable and the current scaling with $\alpha = 3/2$(Eq.\ref{psmain}), which is not very different from $\alpha = 5/3$, is more appropriate. However, in the range of $Re_f\theta_0$ studied in our experiments, the difference between the three scaling exponents is not very large and and often within the margin of experimental uncertainty.  

Since $U^*_{BL} = Re_s/Re_f$, the scaling relation in Eq.\ref{psmain} can also be expressed as 
\begin{equation}
    U^*_{BL} \sim (1-2p)Re_f^{1/2}\theta_0^{3/2}
    \label{ubl:ps}
\end{equation}
The \textit{power scaling} relation in terms of $U^*_{BL}$ plotted in figure\ref{ubl:Pow}. The $Re_f$ dependence of $U^*_{BL}$ in Eq.\ref{ubl:ps} highlights the viscosity dependence of the this regime.
%-----------------------------------
\subsection {Separable scaling regime}
\label{ssr}
When pitched at moderate to large amplitudes ($\theta_2$,$\theta_3$ and $\theta_4$), we argue that the primary resistance is due to pressure drag \citep{quinn2014, moored2019}. The pressure difference across the airfoil $\Delta p \sim U_s^2$. The pressure force acts on a projected area proportional to the pitching amplitude $\theta_0$, resulting in a pressure drag coefficient $\bar C_{D} \sim \theta_0$. In self-propelling state, this drag balances the invisicd reactive thrust  (Eq.\ref{Ct:R}) such that
\begin{equation}
k_s \sim (1-2p)^{-1/2} \theta_0^{-1/2}
\label{ss1}
\end{equation} 
which can be rewritten in terms of Reynolds number
\begin{equation}
Re_s \sim (1-2p)^{1/2}Re_f \theta_0^{1/2}
\label{ssmain}
\end{equation}
The compensated plot for $Re_s$ in figure \ref{Re:Sep} shows that the experimental data collapses well with the predicted scaling relation. We also compare the \textit{separable} scaling relation with the self-propulsion data of \citet{das2022} where a strong amplitude dependence is reported. Note that their data collapses much better with the scaling relation in Eq.\ref{ssmain} than the amplitude-dependent power-scaling reported in their study. Our relation easily captures the amplitude effects without the need to arbitrarily adjust the constants to account for amplitude variations. Furthermore, the form of scaling relations as in Eq.\ref{ss1} also highlights the viscosity independence of the self-propulsion in this regime. 

In this regime, for a given pitching point location, $Re_s\sim Re_f$ but $Re_s\sim \theta_0^{1/2}$ indicating a linear increase in $U_s$ with $f$ and as $\theta_0^{1/2}$. Unlike the \textit{power scaling regime} where the increase in $Re_f$ (or $f$) and $\theta_0$ has the same effect on $Re_s$, in the \textit{separable scaling regime}, the effect of increase in $Re_f$ (or $f$) is much stronger than the increase in $\theta_0$. The \textit{separable scaling regime} highlights the diminishing effect of pitching amplitude on self-propulsion speed. The increase in pressure drag off-sets the gain in propulsive thrust achieved by increasing amplitude.

The identification of this regime and the validation of it's existence through our experimental data is an important and perhaps the most significant contribution of this study. It further emphasizes the prominence of pressure drag as a dominant resistance to self-propulsion and not viscous (enhanced) skin friction at these pitching parameters. Interestingly, for majority of fish species, which propel themselves through caudal fin oscillations, the forward speed (or $Re_s$) is observed to increase linearly with the frequency of oscillation (or $Re_f$) particularly for $Re_s \sim O(10^3)$ \citep{videler1993, gazzola2014}. 

In figure \ref{ubl:Sep} we plot this \textit{separable scaling regime} in terms of $U^*_{BL}$ and the scaling relation is of the form
\begin{equation}
    U^*_{BL} \sim (1-2p)^{1/2}\theta_0^{1/2}
\end{equation} 
Note the absence of the influence of viscosity on $U^*_{BL}$. 
%--------------------------------------------------------------
%
\subsection{Linear scaling regime}
\label{lsr}
\begin{figure}
    \centering
    \begin{subfigure}{0.325\textwidth}
        \includegraphics[width = \textwidth]{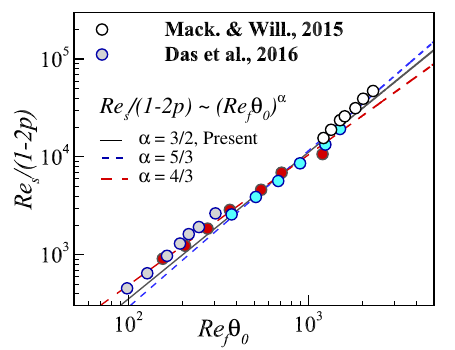}
        \caption{Power Scaling Regime}
        \label{Re:Pow}
    \end{subfigure}
    \begin{subfigure}{0.325\textwidth}
        \includegraphics[width = \textwidth]{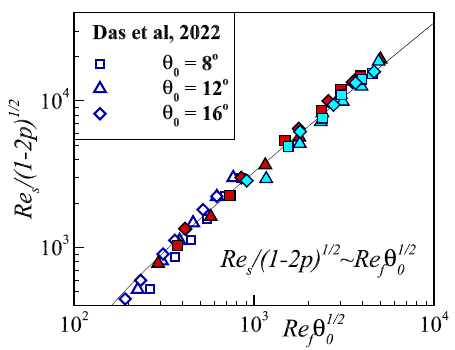}
        \caption{Separable Scaling Regime}
        \label{Re:Sep}
    \end{subfigure}  
    \begin{subfigure}{0.325\textwidth}
        \includegraphics[width = \textwidth]{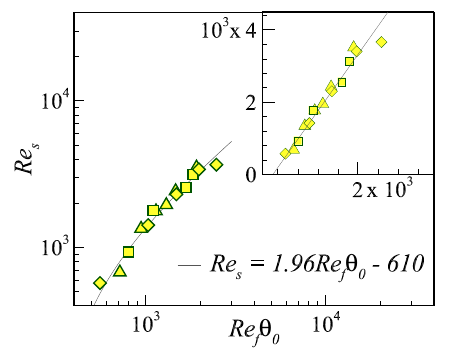}
        \caption{Linear Scaling Regime}
        \label{Re:Lin}
    \end{subfigure}
    \caption{Three scaling regimes: (a)\textit{power scaling regime} for $\theta_1$ and pitching locations P1 and P2 compared with \citet{mackowski2015} and \citet{das2016} for $p = 0.25C$ and $\theta_0 = 2\deg$ and $\theta_0 = 2\deg$ respectively;(b) \textit{separable scaling regime} for $\theta_2,\theta_3$ \&$\theta_4$ and pitching location P1 and P2 and with data points from \citet{das2022} collapsing much better with Eq. \ref{ssmain} (c)\textit{linear scaling regime} when pitched at P3.  c.The inset in the (c) shows the same relation in a linear plot. Legend is same as figure\ref{U_s}}
    \label{Re:sc}
\end{figure}

\begin{figure}
    \centering
    \begin{subfigure}{0.325\textwidth}
        \includegraphics[width = \textwidth]{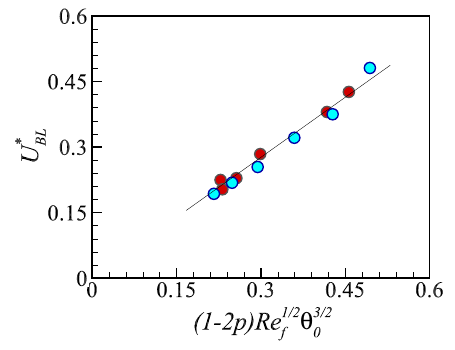}
        \caption{Power Scaling Regime}
        \label{ubl:Pow}
    \end{subfigure}
    \begin{subfigure}{0.325\textwidth}
        \includegraphics[width = \textwidth]{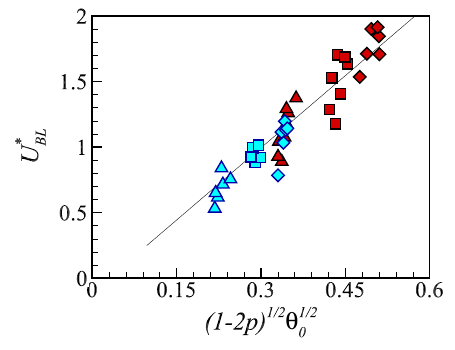}
        \caption{Separable Scaling Regime}
        \label{ubl:Sep}
    \end{subfigure}  
    \begin{subfigure}{0.325\textwidth}
        \includegraphics[width = \textwidth]{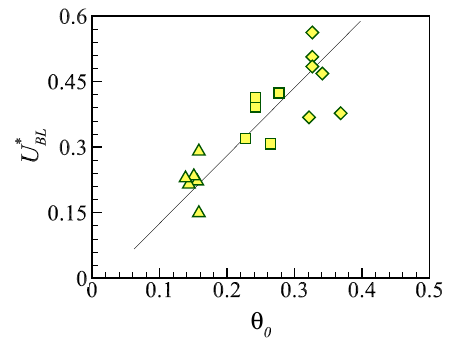}
        \caption{Linear Scaling Regime}
        \label{ubl:Lin}
    \end{subfigure}
    \caption{Scaling regimes represented in terms of normalized speed $U^*_{BL}$. Legend is the same as in figure \ref{U_s}. The solid line represents the best fit.}
    \label{UBL_sc}
\end{figure}

In figure \ref{Re:Lin} and \ref{ubl:Lin} we plot the \textit{linear scaling regime}, in terms of $Re_s$ and $U^*_{BL}$, valid when pitching at P3, $p = 0.48$ (almost near the mid-chord). We observe that $Re_s$ increase linearly with $Re_f\theta_0$  beyond a threshold value and the relation is best captured by the curve $Re_s = 1.96Re_f\theta_0 - 610$. Correspondingly, $U^*_{BL} \sim \theta_0$. The linear relationship reported here is similar to the linear $Re_s - Re_f$ relation in heaving self-propelling flat-plates and elliptical airfoils \citep{vandenberghe2004, spagnolie2010}. 

\begin{figure} 
\centerline{\includegraphics[width = 1\textwidth]{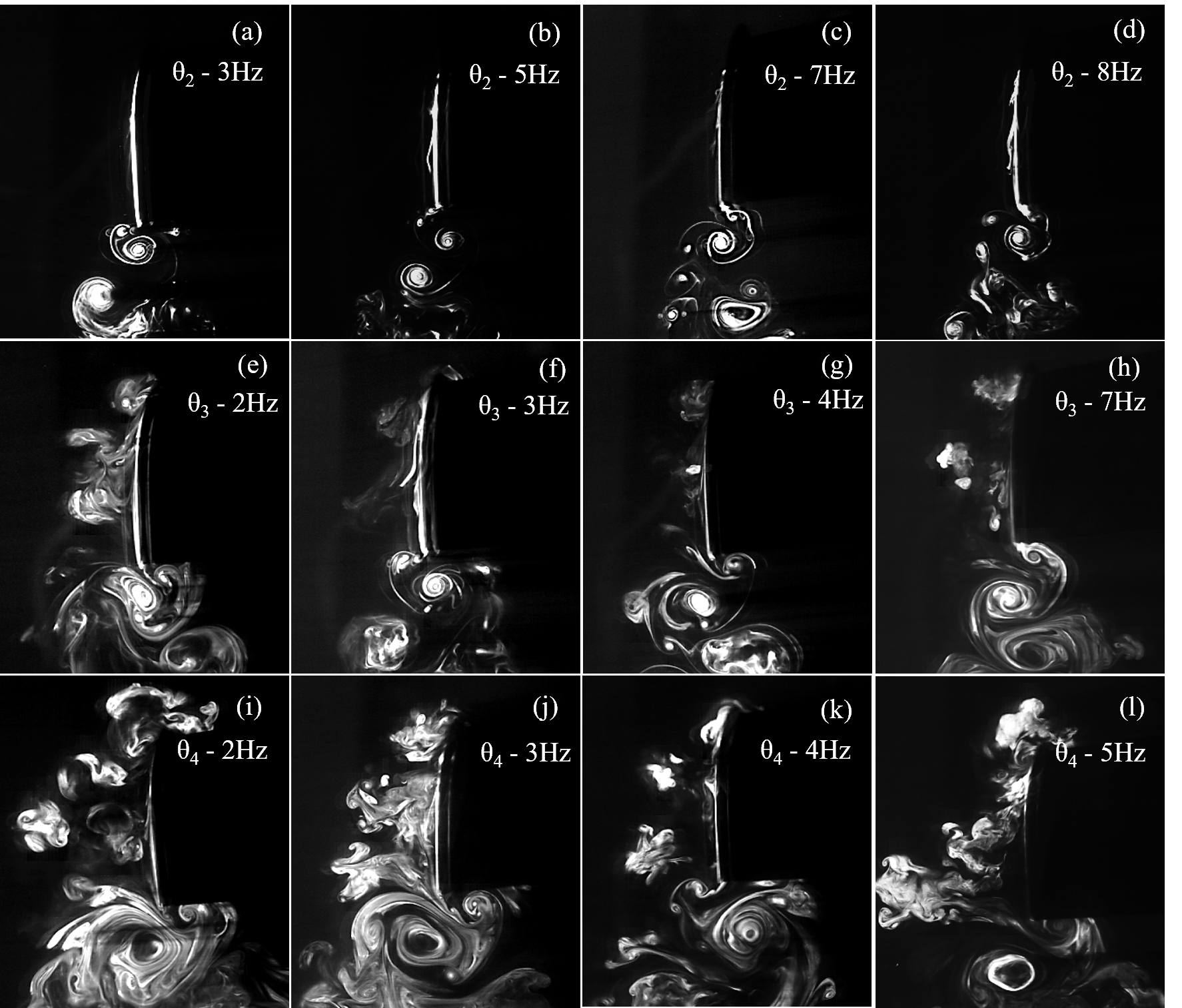}}
\caption{Instantaneous snapshots of flow around the airfoil when pitching at P3 ($p = 0.48$) when the airfoil is at the mean pitching position.}
\label{fig:P3_boundary}
\end{figure}

We discussed earlier (section \ref{thrust_gen_in}) that when pitched at P3 with $p \approx 1/2$, the average invisicd thrust is due to leading edge suction and the total invisicd thrust scales as in Eq.\ref{Ct:LeS}. A linear relation between $Re_s$ and $Re_f\theta_0$ is possible only when the invisicd thrust (due to leading edge suction) is balanced by an \textit{enhanced pressure drag} of the form $\bar C_{D} \sim k_s\theta_0$. The presence of highly unsteady and vortical flow field around the airfoil boundary at P3 could lead to an enhancement in pressure drag. Figure \ref{fig:P3_boundary} presents the snapshots of the instantaneous flow field  around the airfoil at mid-span when pitched at P3 obtained using LIF. Notice the presences of large flow structures along the airfoil boundary and leading edge. Such flow patterns along the airfoil boundary are visible only at P3, whereas for other pitching locations, the dye is smoothly swept downstream and vortices are shed in the wake. Quantitative information about the flow field and the unsteady boundary layer on the airfoil is essential to fully understand the drag mechanism. Nevertheless, the proposed scaling relation remains robust and represents a significant contribution to this field of study. 

%-----------------------------------------------------------

%\subsection{Scaling relations in terms of $U^*_{BL}$ and $U^*_{AL}$}

%----------------------------------------------------------------------------
\section{Vortex shedding in the wake}
\label{wake}
\begin{figure} 
\centerline{\includegraphics[width = 1\textwidth]{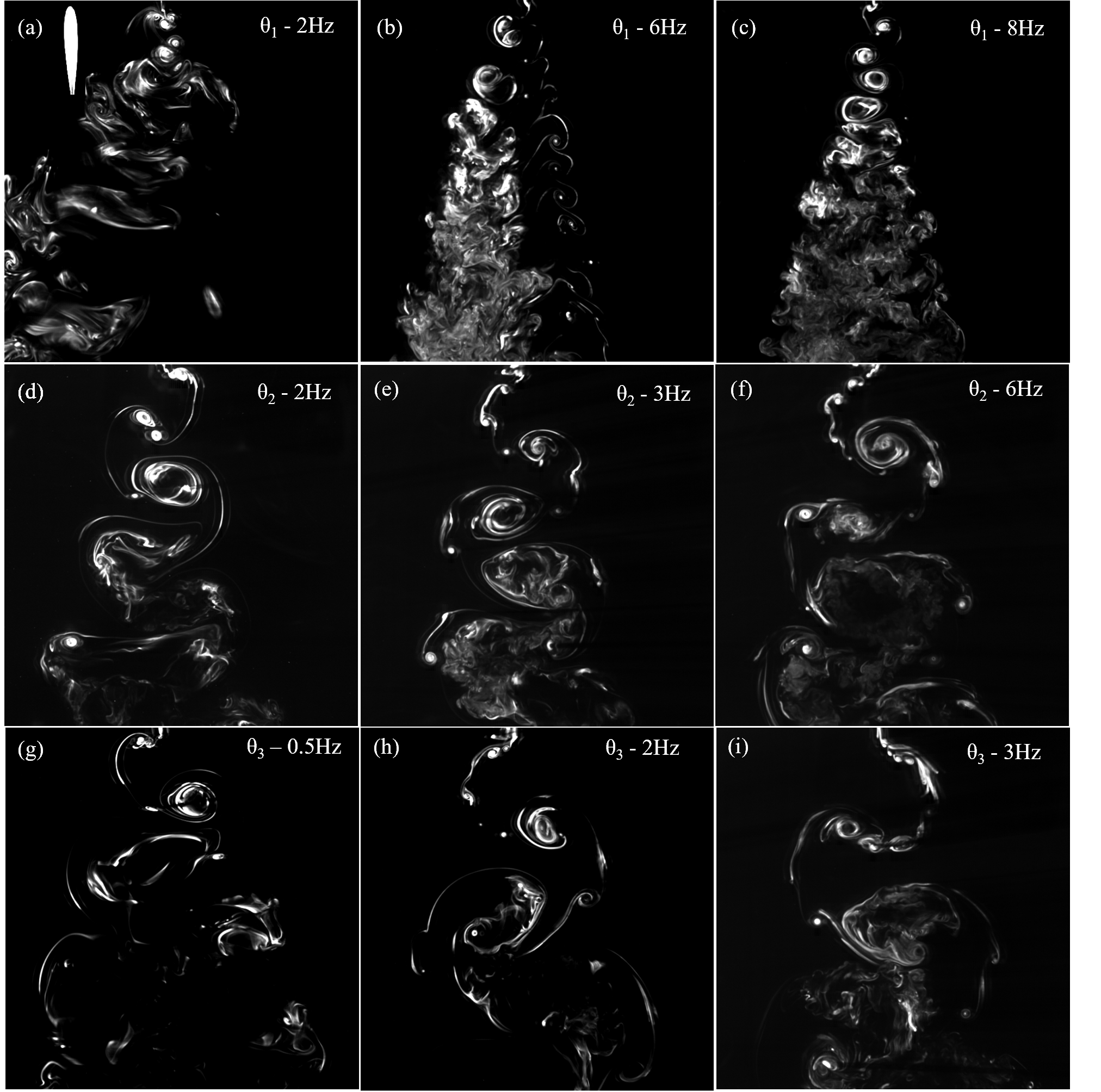}}
\caption{The wake vortices observed for pitching point P1. From the top left, we see (a) Deflected Vortex Pair (DVP), (b) DVP with a single vortex (DVP -S), (c) Reverse von Karman (RvK), (d) RvK, (e) RvK with smaller vortex; (f) RvK with two smaller vortices; (g) RvK; (h) and (i) RvK-C with multiple smaller vortices coalescing into a larger one. The airfoil image in (a) provides a reference scale}
\label{fig:P1_vortex}
\end{figure}

\begin{figure} 
\centerline{\includegraphics[width = 1\textwidth]{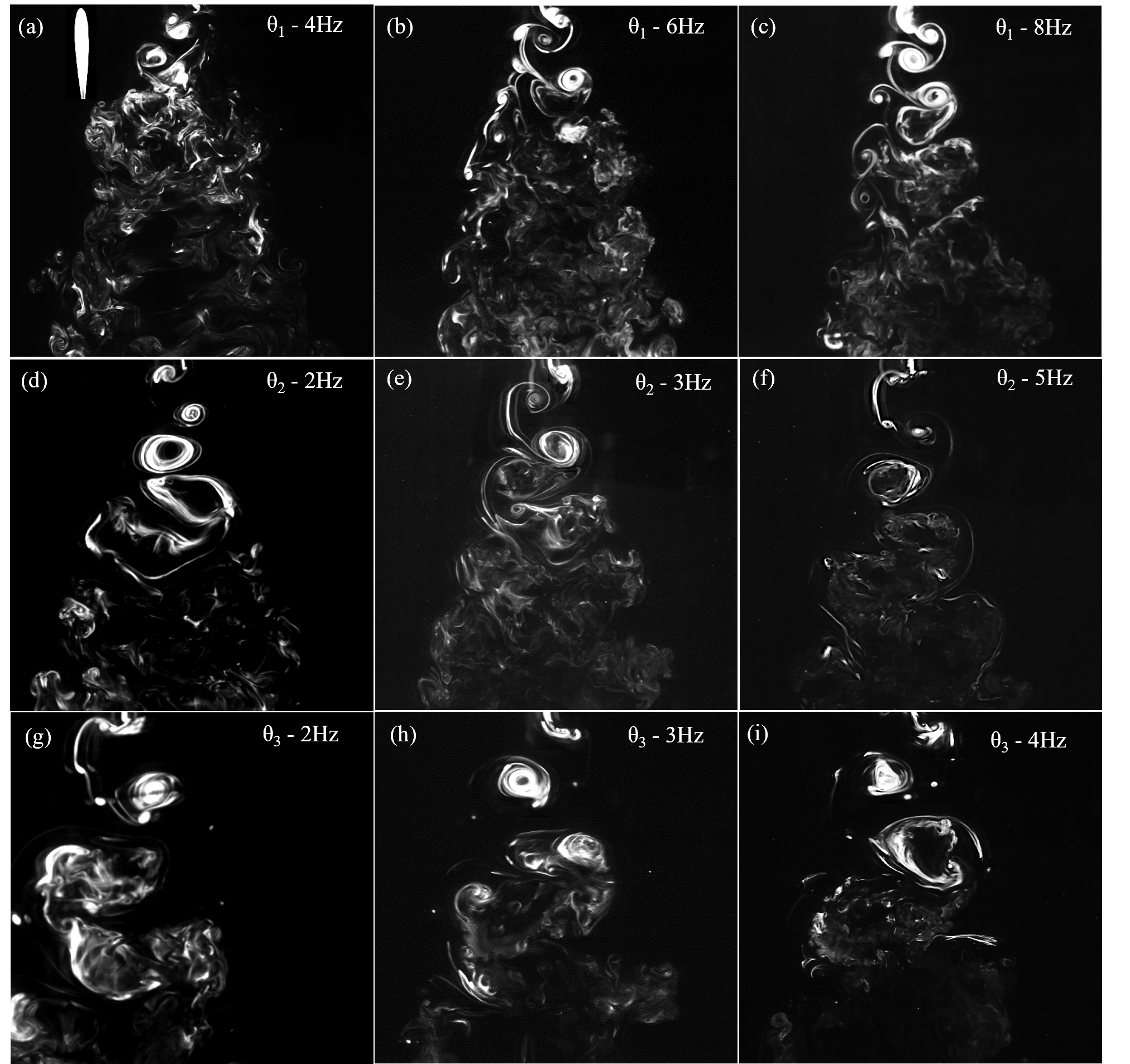}}
\caption{The wake vortices observed for pitching point P2. Deflected vortex pair (DVP) in (a); Deflected vortex pair with a single vortex (DVP-S) in (b). Reverse von Karman (RvK) like in (c) – (f) and RvK coalescing (RvK-C) in (g)-(i). The airfoil image in (a) provides a reference scale}
\label{fig:P2_vortex}
\end{figure}

\begin{figure} 
\centerline{\includegraphics[width = 1\textwidth]{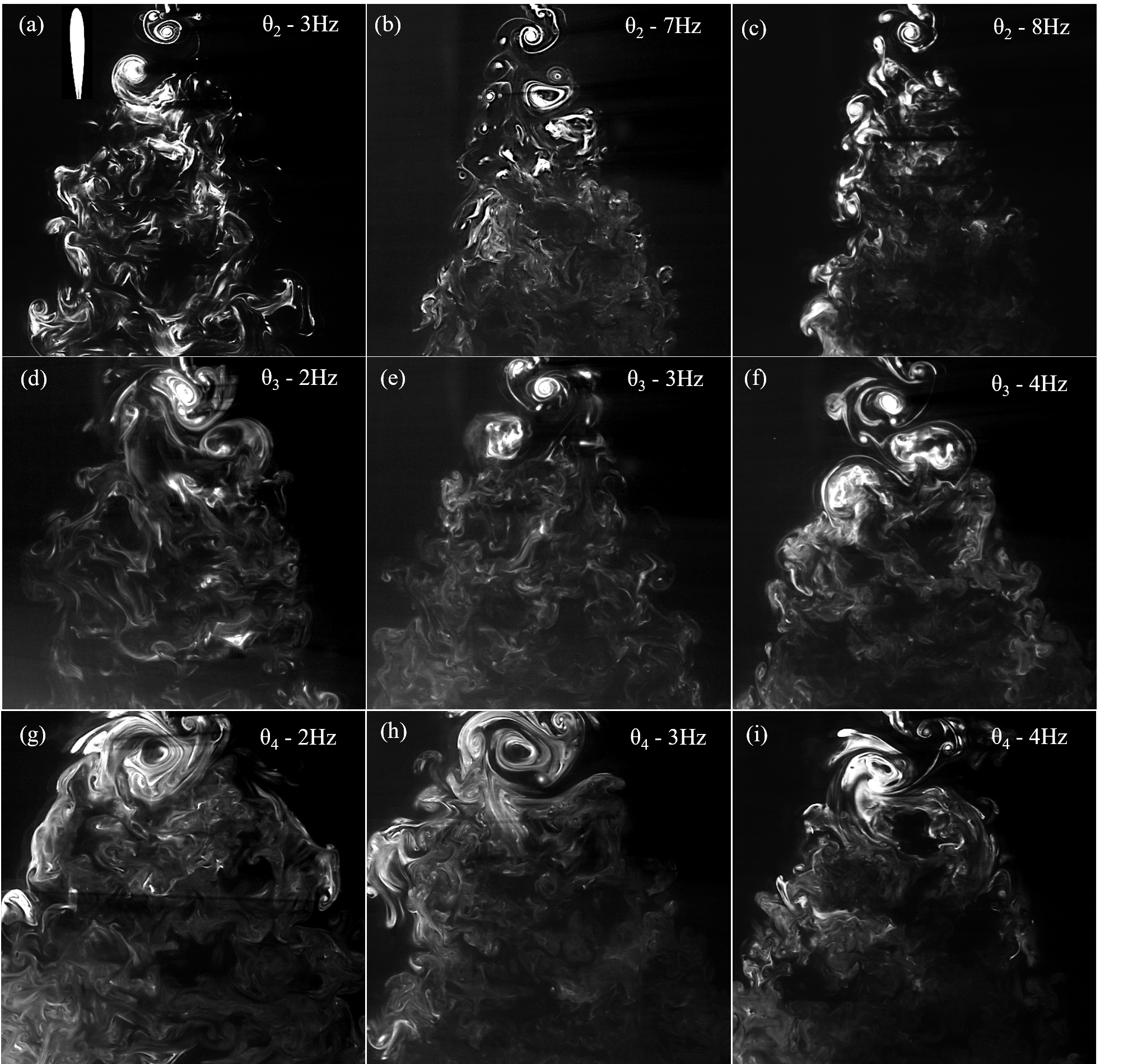}}
\caption{The wake vortices observed for pitching point P2. Deflected vortex pair (DVP) in (a)-(f). Unsteady reverse von Karman (RvK) like pattern in (g)-(i). The airfoil image in (a) provides a reference scale}
\label{fig:P3_vortex}
\end{figure}

Figures \ref{fig:P1_vortex}, \ref{fig:P2_vortex} and \ref{fig:P3_vortex} show the snapshots of vortex patterns in the wake through the dye visualisation when pitching at P1, P2, and P3, respectively, representing all of the observed patterns. In general, the wake of the self-propelling rigid airfoil is characterized by a wide and unsteady flow field. Distinct vortices are identifiable only in the immediate downstream region of the airfoil, about a chord length. Further downstream, the wake is wide and unsteady but distinct vortices are no longer identifiable. This observation is in stark contrast with the earlier studies where distinct vortices persist further downstream even at drag-thrust cross over (self-propulsion). However, these experiments use end plates which suppress the 3-dimensional evolution of the flow (see for example \citet{mackowski2015}). In our experiments, the interaction between the stream-wise vorticity with the span-wise vorticity is expected to be much stronger due to the absence of such an end plate. This interaction is well captured and explained by the vortex evolution model of \citet{ellenrieder_2003,buchholz_2008} for pitching airfoils of finite amplitudes. In the immediate downstream region of the airfoil, the span-wise vorticity is strong and the flow around the airfoil is largely two-dimensional except at the edges with a weak stream-wise vorticity. The spanwise vorticity along with the stream-wise vorticity forms a horse-shoe which is stretched as it travels downstream (or the airfoil travels upstream). See figure 4 in \citet{ellenrieder_2003} and figure 24 in \citet{buchholz_2008} for sketches. The span-wise vorticity is pulled away from the airfoil center line as it continues to grow weaker while the stream-wise vorticity grows stronger. A couple of chord-lengths downstream the horse-shoe structures become even more convoluted and inter meshed. At this point, the vorticity distribution is concentrated more around the mid-span and away from center-line, visible as a wide and unsteady wake as reported here.

Nevertheless, the distinct vortex pattern in the immediate vicinity of the airfoil is directly correlated to $U^*_{BL}$ and therefore the near-wake vortex pattern can be an identifier of the scaling regime of the self-propulsion to a certain extent. We classify vortex patterns in the near wake region as Deflected Vortex Pair (DVP, eg. figure\ref{fig:P1_vortex}a), Reverse von-Karman (RvK, eg. figure \ref{fig:P2_vortex}d) vortices and RvK-Coalasence (RvK-C, eg. figure \ref{fig:P2_vortex}g). DVPs were also reported by \citet{das2016} and Godoy-Diana et. al (2008) when the airfoil still experienced a net drag. In self-propulsion or drag-thrust transition, the RvK is the most reported pattern. Here we see that DVP can also occur in self-propelling airfoils for lower $U^*_{BL}$. 

\begin{figure} 
\centerline{\includegraphics[width = 0.45\textwidth]{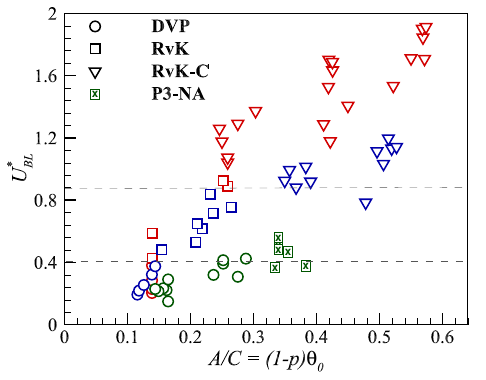}}
\caption{A map illustrating the largely three different wake patterns observed for the self-propelling rigid airfoil as a function of $U^*_{BL}$ and the trailing edge excursion $A/C = (1-p)\theta_0$. The red symbols are for P1, blue for P2 and green for P3.}
\label{fig:map_vortex}
\end{figure}

We map these vortex patterns in a $U^*_{BL} - A/C$ map in figure \ref{fig:map_vortex}. In general, a vortex pair is shed for every oscillation cycle. The occurrence of different patterns in the wake can be primarily attributed to the spatial separation between the two shed vortices which directly depends on the longitudinal distance traveled by the airfoil $\sim U^*_{BL}$. For $U^*_{BL} < 0.4$, DVP is the prominent pattern in the wake. Here, the spatial separation between the shed vortices is smaller, resulting in a stronger interaction between the vortices and the formation of DVP. These patterns can be seen in figure \ref{fig:P1_vortex} (a) and figure \ref{fig:P2_vortex} (a-b). Note that the direction of deflection is rather arbitrary for a given trial. As the speed increases (with frequency), we observe a gradual transition from DVP to DVP with a single vortex (DVP-S) and then eventually to the Reverse von Kármán (RvK) pattern (figure \ref{fig:P1_vortex}(b-c)).  For $0.4<U^*_{BL}< 0.8$, RvK pattern is present. When pitching at higher amplitudes ($\theta_2$-$\theta_4$), we first observe a RvK (figure \ref{fig:P1_vortex}(d), \ref{fig:P2_vortex}(d,e)). With a further increase in speed ($U^*_{BL} > 0.8$), a larger vortex along with several smaller vortices are shed every half cycle. The smaller vortices eventually merge into a larger vortex. The number of these smaller vortices increases with the increase in speed (see, for example, figure \ref{fig:P1_vortex}(d,e,f)), resulting in RvK-C. These larger merged vortices still maintain an overall RvK configuration. Note that in the vortex map (figure \ref{fig:map_vortex}) we consider DVP-S within the broader category of the DVP pattern itself. When pitching at P3, the vortex patterns are closest to the DVP pattern (figure \ref{fig:P3_vortex}) owing to lower $U^*_{BL}$, although the flow field is highly unsteady in the wake and even around the airfoil. For P3 at $\theta_4$, the wake pattern cannot be classified into any of the above groups and is labeled as NA.   

%----------------------------------------------------------------------------
\section{Conclusions}
\label{concl}
The impact of imposed kinematics on self-propulsion becomes evident through the existence of three distinct regimes of self-propulsion, each characterized by a unique relationship between $Re_s$, $Re_{f}$, $\theta_0$ and $p$. The three scaling regimes, i.e. {\it power, separable}, and {\it linear} regimes highlight the invisicd nature of thrust generation and the different drag mechanisms namely the enhanced skin friction, the pressure and possibly, an enhanced pressure drag, respectively. As such, the scaling relations in Eq. \ref{psmain} and \ref{ssmain} and $Re_s \sim Re_f\theta_0$ capture the self-propulsion characteristics across all the pitching parameters. Different vortex patterns are observed and we highlight the dependence of these patterns on $U^*_{BL}$ which dictates the spatial separation between shed vortex cores. 

The present study consolidates our understanding of self-propulsion in oscillating airfoils and provides the much needed extensive experimental groundwork on this subject.We have specifically examined the kinematic or the 'speed' aspect in exploring the self-propulsion in oscillating airfoils with significant rigor. However, a crucial question of effect of imposed kinematics on efficiency remains unanswered and forms the basis of future work.
%-------------------------------------------------------------------------------------------------------------------------
\backsection[Supplementary data]{There is no supplementary data.}

\backsection[Acknowledgments]{The authors acknowledge the discussions with Dr. Anil Das, Prof. Ratnesh Shukla, and Prof. Raghuram Govardhan on the subject matter of this article.}

\backsection[Funding]{This study was partially funded by the Naval Research Board. The first author received a subsistence fellowship provided by FERCC during the initial preparation of this manuscript.}

\backsection[Declaration of interests]{The authors report no conflict of interest.}

\backsection[Data availability statement]{The data that support the findings of this study can be shared on a reasonable request.}

\backsection[Author ORCIDs]{Rakshita Joshi, https://orcid.org/0000-0001-8831-8790; Jaywant Arakeri https://orcid.org/0009-0001-2575-0604}

\backsection[Note]{The given name of the first author, as per official records, is Rakshitha Ulhas Joshi. The author publishes under the name Rakshita Joshi.}

%--------------------------------------------------------------------------------------------------------------------------
\appendix

\section{Inviscid inertial thrust scaling}
\label{invisicdthrust}

Here, we revisit the linear theory of \citet{garrick1936} and \citet{theodorsen1935} in detail to determine the expressions for the invisicd thrust coefficients. Consider a flat plate airfoil of chord length $C$ pitching about a point located at a distance $pC$ from the leading edge in a uniform free-stream. Recall that $p = 0$ at the airfoil leading edge and $p = 1$ at the trailing edge. A sinusoidal pitching motion of the airfoil can be represented in a complex form such that the instantaneous angle of the airfoil with respect to the free-stream is given by
\begin{equation}
    \theta = \theta_0 e^{i\omega t}
\end{equation}
where $\omega = 2\pi f$ is the angular frequency. $k = \pi fC/U$ is the reduced frequency and $U$ is the free-stream velocity. 

When the airfoil carries out it's motion the invisicd forces acting on it are due to the fluid reaction ($P_R$), circulatory forces due to instantaneous non-zero angle of attack and vorticity in the wake ($P_C$), and the leading edge suction ($P_{Les}$). Here, the wake is modeled as a thin vortex sheet of strength $\gamma_w(x)$ across which there is a jump in the velocity \citep{theodorsen1935, garrick1936}. Although this wake model is vastly different from the observed wake features, we establish that in the limit $k \rightarrow \infty$, the dominant forces on the airfoil are due to fluid reaction and leading edge suction. Therefore, the inaccuracy in the nature and form of wake model considered here has negligible consequences. Further, 
\begin{equation}
P_{R} = -\pi \rho b^2 (U \dot{\theta} - ba \ddot{\theta})
\label{eqnPR}
\end{equation}
with $\dot{\theta}$ and $\ddot{\theta}$ representing the first and second derivative of $\theta$ with time, and

\begin{equation}
P_{C} = -2\pi\rho U^2bC(k)Q(t)
\label{eqnPC}
\end{equation}
$\mathcal{C}(k) = F(k) + iG(k)$ is the Theodersen function of $k$ such that
\begin{align}
F = \frac{J_1 (J_1 + Y_0) + Y_1(Y_1 - J_0)}{(J_1+Y_0)^2 +(Y_1 - J_0)^2} 
\label{eqnF}\\ 
G = -\frac{Y_1 Y_0 +J_1 J_0}{(J_1+Y_0)^2 +(Y_1 - J_0)^2}
\label{eqnG}
\end{align}
$J_0(k)$, $J_1(k)$, $Y_0(k)$ and $Y_1(k)$ are the bessel functions of the first and the second kind respectively \citep{bessel2010} and
\begin{equation}
    Q(t) = -U\theta + \frac{C}{2}\left(2p -\frac{3}{2}\right)\dot{\theta}
\end{equation}
At any instant, the total propulsive force generated by a sinusoidally pitching airfoil results from the linear superposition of the different thrust components such that
\begin{equation}
    P_x = \Im(P_R)\Im(\theta) + \Im(P_C)\Im(\theta) + P_{LeS}
    \label{Px}
\end{equation}
The leading edge suction
\begin{align}
    P_{LeS} = \pi\rho\frac{C}{2}[\Im(S)]^2;\\
    S = \frac{1}{\sqrt{2}} \left(2\mathcal{C}(k)Q(t) + \frac{C}{2}\dot{\theta}\right)
\end{align}
The thrust forces given in Eq.\ref{Px} can be normalized by $\rho U^2C/2$ to obtain the respective thrust coefficients. 

The instantaneous net thrust coefficient for a pitching airfoil can be resolved into an average and an unsteady component such that
\begin{equation}
C_T = \bar{C}_T + C_{T,0}\sin \left(2\omega t - 2\phi \right)
\end{equation}
Here, $\bar{C}_T$ is the average thrust coefficient, $C_{T,0}$ is the amplitude of oscillatory component of the thrust and $\phi$ is the phase lag with the respect to the pitching motion. As discussed earlier in section \ref{thrust_gen_in}, the $C_{T,0}$ and $\phi$ values predicted by \citet{garrick1936} match very well with the amplitude and phase of unsteady force measurements reported by \citet{mackowski2015} and the net thrust can be determined by Eq. \ref{thrust_net}. The total average invisicd thrust coefficients would then be a linear superposition of the individual average coefficients with
\begin{equation}
\bar{C}_{T,in} = \bar{C}_{T,R}+ \bar{C}_{T,C} + \bar{C}_{T,LeS}
\label{thrust}
\end{equation}
Following \citet{garrick1936}, we get the following expressions for all the components and the total average thrust coefficients.
\begin{align}
\bar{C}_{T,R} & = \pi \theta_0^2 \frac{(1-2p)}{2}k^2 \label{reac} \\
\bar{C}_{T,C} & = \pi \theta_0^2 \left(- F + k\left(\frac{3}{2}-2p\right)G\right) \label{cir} \\
\bar{C}_{T,LeS}& = \pi \theta_0^2 S_0 \label{les} \\
\bar{C}_T & = \pi \theta_0^2 \left(\frac{1-2p}{2}k^2 + T_0\right) \label{total}
\end{align}
Where, 
\begin{align*}
S_0 & =\left(F - k\left(\frac{3}{2}-2p\right)G\right)^2 +\left(G + k\left(\frac{3}{2}-2p\right)F - \frac{k}{2}\right)^2 \\
T_0 & = -F + k\left(\frac{3}{2}-2p\right)G + \left(F - k\left(\frac{3}{2}-2p\right)G\right)^2 + \left(G + k\left(\frac{3}{2}-2p\right)F - \frac{k}{2}\right)^2
\end{align*}
Note that the expression for total thrust (Eq.34) in \citet{garrick1936} is incorrect and contains an algebraic error. The corrected version is present in \citet{garrick_1957} although without any reference to the error in \citet{garrick1936}. This discrepency was earlier noted and pointed out by \citet{jones_1997}. 
\subsection{Asymptotic behaviour} 
In the limit $k \rightarrow \infty$, the following relations hold for bessel functions of the first and second kind \citep{bessel2010}
\begin{align*}
J_0 \approx \left(\frac{2}{\pi k}\right)^{1/2}\left\{\cos\left(k -\frac{\pi}{4}\right)+ \frac{1}{8k}\sin\left(k-\frac{\pi}{4}\right)+ O(k^{-2})\right\}\\
J_1 \approx \left(\frac{2}{\pi k}\right)^{1/2}\left\{\cos\left(k -\frac{3\pi}{4}\right) - \frac{3}{8k}\sin\left(k-\frac{3\pi}{4}\right)+ O(k^{-2})\right\}\\
\\
Y_0 \approx \left(\frac{2}{\pi k}\right)^{1/2}\left\{\sin\left(k -\frac{\pi}{4}\right) - \frac{1}{8k}\cos\left(k-\frac{\pi}{4}\right)+ O(k^{-2})\right\}\\
Y_1 \approx \left(\frac{2}{\pi k}\right)^{1/2}\left\{\sin\left(k -\frac{3\pi}{4}\right) + \frac{3}{8k}\cos\left(k-\frac{3\pi}{4}\right)+ O(k^{-2})\right\}
\end{align*}
Consequently, $F \sim 1/2$ and $G \sim 0$ as $k \rightarrow \infty$. Therefore, $\bar{C}_{T,C} \approx 0$ while
\begin{equation}
    \bar{C}_R = \pi \theta_0^2\frac{k^2}{2}\left(1-2p\right);
%    \label{Ct:Reac}
\end{equation}
\begin{equation}
    \bar{C}_{LeS} \approx \pi \theta_0^2 \frac{k^2}{4} \left( 2p - \frac{1}{2} \right)^2
    \label{Ct:Les}
\end{equation}
and the total invisicd thrust
\begin{equation}
    \bar{C}_{T,in} \approx \pi\theta_0^2\frac{k^2}{4}\left(\frac{3}{2} -2p\right)^2
\end{equation}

%-------------------------------------------------------------------------------------------------------------------------
\bibliographystyle{jfm}
\bibliography{Scaling_References}

\end{document}